# DatChain - Blockchain implementation in Data transfer for IoT Devices

Om Rajput

Department of Computer Science and Engineering

SRM Institute of Science and Technology

or6857@srmist.edu.in.

Suyash Nigam

Department of Computer Science and Engineering

SRM Institute of Science and Technology

sn7821@srmist.edu.in.

Dr. M.J. Chowdhury

Department of Computer Science and Engineering

La Trobe University

M.Chowdhury@latrobe.edu.au

Dr. Kayalvizhi Jayavel

Department of Computer Science and Engineering

SRM Institute of Science and Technology

kayalvij@srmist.edu.in.

**Abstract** - Currently, the IoT ecosystem is comprised of fully connected smart devices that exchange data to provide more automated, precise, and fast decisions. This idealised situation can only be accomplished if a system for data transactions is processed efficiently and security is ensured with high scalability and practicability. The integrity of data must be maintained during the exchange or transfer of data between entities.

We propose to make a application called DatChain that responds to the above situation. The application stores data sensed by the Iot sensors in the backend after encrypting it and when the data is required for any purpose it can be exchanged using a suitable blockchain network that can keepup with the transfer rate even at high traffic in a secure environment.

## I. INTRODUCTION

The lot market is expanding rapidly and will likely undergo multiple transformational cycles. Likewise, concerns pertaining to privacy, security, and trust will evolve and transform as the situation changes. IoT vendors must be able to handle the increased throughput and develop workable strategies.

Typically, IoT applications for data transfer are developed using the client-server model, in which Internet-connected smart devices are linked to cloud-based servers. Using the underlying communication network, the cloud receives the data collected by the smart devices immersed in the physical environment. Consequently, the majority of computational data processing occurs in the cloud. Regardless, the traditional cloud-centric conceptualisation may result in a design bottleneck due to the daily accumulation of more and more smart devices. By 2030, it is anticipated that more than fifty billion devices will be connected. In addition, the global market for IoT solutions for end users reached 100 billion dollars in 2017 and is anticipated to reach $1.6 trillion by 2025.

Under the perspective of a layered architecture, it is therefore reasonable to consider transferring a portion of the cloud function located in upper layers to intermediate layers closer to the layer of smart devices. Notably, edge computing and fog computing will be utilised for this purpose. In this situation, there must be gateways in intermediate layers to process the data traffic generated by smart devices. These gateways may interact with one another and the cloud itself, if necessary. This new distributed and decentralised paradigm is hoped to be more efficient than the traditional cloud-centric paradigm. How to construct a system database that enables efficient and secure transaction processing for smart devices is one of the major concerns in the scenario described above. Effectiveness correlates to rapid data processing. Also known as the CIA triad, security relates to confidentiality (i.e., available/disclosed only to authorised users), integrity (i.e., reliable and immutable), and availability (i.e., available when needed).

To meet these requirements, academia and industry have considered deploying Blockchain, a distributed ledger technology (DLT). In order to be safe, scalable, and efficient, the traditional centralised IoT cloud systems will be converted to blockchain-based systems by 2025, according to a report from a survey conducted by IBM.

Nevertheless, despite the successful case studies and deployments already presented in the literature, there is still a need to confirm the Blockchain's efficiency when used in the IoT ecosystem, as the corresponding applications are numerous, execute in parallel, and possess quite intrinsic characteristics, such as transaction rate, number of communicating devices, and business logic. Moreover, the model must also support the business needs and feasibility.

## II. GROUNDWORK

### 1. IoT Ecosystem

By Definition, the IoT ecosystem is a connection of various kind of devices that sense and analyze the data and communicates with each other over the networks. The Internet of Things ecosystem includes all the components that enable businesses, governments, and consumers to connect to their IoT devices. The ecosystem includes remotes, dashboards, networks, gateways, analytics, data storage, and security.

In the IoT ecosystem, the user uses smart devices such as smartphones, tablets, sensors, etc. to send commands or requests to devices for information over the networks. The device responds and performs the command to send information back to the user through networks after being analyzed.

### 2. Blockchain Fundamentals

A blockchain is a shared, distributed database or ledger among the nodes of a computer network. Blockchain technology is a collection of concepts, such as peer-to-peer protocols, hashing algorithms, and distributed consensus algorithms. Each of these mechanisms is described in the following section.

### 2.1 Peer-to-peer Protocols

A peer-to-peer network has contributed to the establishment of Blockchain, which is a decentralised consensus network. A peer in a peer-to-peer network is also referred to as a node. No peer is superior to others in a peer-to-peer network, and all nodes share the responsibility of providing the necessary network services. Without a hierarchy, central authority, or primary server, the peer-to-peer network is completely decentralised due to its flat topology. A consensus mechanism is used in a peer-to-peer network to ensure that the block is valid before it is recorded on the ledger. The entire network will have a copy of the updated ledger once the block has been recorded on the ledger. All nodes in the peer-to-peer network have access to data and form an autonomous network to distribute and share data. Consequently, the blockchain becomes the only source of truth.

### 2.2 Hashing Algorithm

The blocks within a blockchain are interconnected. This creates a block chain, which is what the name "blockchain" refers to. The first block is called "genesis." Each block has a block index, the hash value of the previous block (also known as the parent block), a timestamp, the block's hash value, and a nonce, a random integer used to verify the hash according to the network's protocols. In addition to the transactions, each block also contains a hash composed of the block's data and the hash of the block that preceded it in the chain.

Nofer et al.highlighted that the hash values are unique, and if any change is made to a block in the chain, the respective hash value would be changed immediately.

To comprehend digital identities, one must comprehend the operation of cryptographic hashing. If an attacker attempts malicious usage by modifying a block's data, the associated hash must be updated. Therefore, if an attacker wishes to modify a single hash, all hashes in the chain between the tampered block and the most recent block must also be modified. However, a successful attack cannot be executed by modifying a single blockchain or ledger on a single node until the majority of ledgers have been replicated. It is quite difficult to duplicate more than fifty percent of the nodes in a short period of time. Consequently, it is believed that modifying the data in a blockchain is nearly impossible. This hash is the key element in a blockchain that assures data integrity.Kaushik et al. and Selmanovic revealed that a good hashing algorithm needs to fulfil the following requirements and SHA256 is the most common hashing algorithm in blockchain platforms.

## 2.3 Consensus Algorithm

By adopting the Byzantine Generals' problem, Lamport et al. suggested a fault tolerance technique to resolve dependability difficulties of a fault-tolerant system. A modification of the Byzantine Generals' issue assisted in achieving consensus among the unreliable nodes of the blockchain network. Consensus on the blockchain refers to a series of procedures including the approval and confirmation of a transaction or set of transactions via consensus algorithms. Proof of Work (PoW) or its derivative consensus algorithms are utilised by the majority of public blockchains to ensure that transactions cannot be tampered with. PoW is a superior security strategy provided that no single miner controls more than fifty percent of the network's hashing power. However, a PoW process requires a large amount of computer power to do numerous calculations, resulting in a high energy need. Bitcoin network uses Proof-of-Work consensus process; as a result of bitcoin mining operations and increasing rivalry for the highest hash power, it consumes approximately 0.33 percent of the world's electricity. As all nodes share a copy of the ledger, once a transaction or set of transactions are added as a block to the ledger, all ledgers reflect this change. The consensus mechanism enables the near-instantaneous update of all ledger copies. There are dozens of consensus algorithms other than PoW, which are aiming to solve mainly three issues: scalability, security, and decentralisation. A good consensus algorithm should address all three factors. However, based on the requirement, consensus algorithms can be more specialised in one or two factors. Most commonly used consensus algorithms are Proof of Work (PoW), Proof of Stake (PoS), Practical Byzantine Fault Tolerance (PBFT), Delegated Proof of Stake (DPoS), Proof of Importance (PoI), Ripple Protocol Consensus Algorithm (RPCA), Stellar Consensus Protocol, and Byzantine algorithm based Tendermint. These are discussed below in detail. Few other known consensus algorithms are Proof of Authority, Proof of Burn, Proof of Activity, Proof of Capacity, Simplified Byzantine Fault Tolerance, Federated Byzantine Agreement, Zero-Knowledge Proof and so on. Some of the common consensus algorithms are mentioned discussed below.

- PoW is the most prevalent consensus algorithm, controlling over 75% of bitcoin market capital. PoW is an algorithm for open consensus. Calculating the hash value of a block with the required amount of leading zeros by altering the nonce (a random number). This procedure is known as mining, and it is an extremely energy-intensive procedure. Once a miner has discovered a valid nonce, it can be disseminated to other nodes for verification.
- PoS was developed as an energy-saving alternative to PoW, and it is a technique with moderate energy consumption. PoS is an algorithm for open consensus. The theory underlying Proof-of-Stake is that there is a low danger of network attack from those with a larger stake. PoS is mostly determined by the age of the coin. The system calculates coin days by multiplying the number of coins held by the number of days held, and then calculates the stake size to determine which node will create the next block. The Cardano cryptocurrency uses the Proof-of-Stake consensus algorithm.
- DPoS has representatives and employs age-based stakes for coins. DPoS is an algorithm for open consensus. The primary distinction between the PoS and DPoS is that the latter uses representative democracy. In DPoS, stakeholders elect delegates to generate and validate blocks on their behalf. The fact that fewer nodes are engaged in the generation of blocks improves the network's performance. DPoS is a consensus protocol with a low to moderate energy footprint.
- RPCA employs a specific voting system with a single or multiple rounds, until all transactions achieve an approval rate of at least 80%. When all nodes acquire at least 80% affirmative votes, it will be recorded in the public ledger. RPCA is both an open consensus algorithm and a method with minimal energy usage.
- PBFT is the consensus protocol utilised by Hyperledger Fabric for permissioned networks. The PBFT method consists of three steps, including pre-prepared, prepared, and commit. At each phase, there is a voting mechanism, and it is important to collect 2/3 votes from all nodes before moving to the next level. After the completion of three steps with more than 2/3 votes, the system will allow adding next block to the ledger. PBFT is a permissioned consensus algorithm and one of the methods with the lowest energy consumption.

3. IOTA Smart Contracts

ISC is integrated into IOTA1's distributed ledger technology (DLT). ISC is a framework that adds many programmable ledgers as layer 2 on top of the IOTA DLT's layer 1 (L1) base protocol (L2). Consequently, all chains are anchored on the L1 IOTA Ledger, establishing a multichain ecosystem. Each chain is a separate blockchain containing smart contracts. These contracts are fully composable and are functionally comparable to smart contracts on Ethereum. The ISC ecosystem also offers high throughput, scalability, and trustless composability of smart contracts via L1 between different chains. The ISC can theoretically scale to hundreds of thousands of smart contract transactions per second as a distributed and sharded multi-chain DLT that comprises both L1 and L2.

## III. LITERATURE REVIEW

This research paper's study was heavily influenced by a small number of previously published papers. This paper is a comparative study and therefore draws conclusions based on the research conducted by other scholars.

**A.** Towards Blockchain for Suitable Efficiency and Data Integrity of IoT Ecosystem Transactions

This research was conducted by carlo Kleber, Da Silva Rodrigues, Vladimir Rocha in the year 2021 and was aimed at Iot Efficiency with data integrity

Processing the transactions made by smart devices and the stored data integrity. Focuses on how Blockchain technology may meet IoT Efficiency requirements.

The article analyzes the effectiveness of deploying the Blockchain technology in the implementation of the IoT ecosystem database. To this end, we assess the processing efficiency of transactions originated by smart devices and the storeddata integrity.

Final results show that the Blockchain technology may meet IoT efficiency requirements, besides providing adequate data integrity. Lastly, general conclusions and avenues for further research close this article[1].

**B.** TrailChain: Traceability of data ownership across blockchain-enabled multiple marketplaces

This study has been conducted by Pooja Gupta, Volkan Dedeoglu, Salil S. Kanhere and Raja Jurdak did an extremely well to achieve its objectives.

These data marketplaces are susceptible to a variety of threats, including unauthorised data redistribution/resale, data tampering, dishonest data ownership claims, and the trade of fake data.

To address the aforementioned issues, they propose TrailChain, a novel blockchain framework that uses watermarking to generate a trusted trade trail for tracking data ownership across multiple decentralised marketplaces. Our solution includes detection mechanisms for the unauthorised reselling of data within and across marketplaces.

In addition, they propose a fair resell payment sharing scheme that ensures data owners receive a portion of the resell revenue from authorised reselling. They present an Ethereum-based prototype implementation of the system. They conduct exhaustive simulations to demonstrate TrailChain's viability by benchmarking execution gas costs, execution time, latency, and throughput[2].

**C.** Resilience of IOTA Consensus

This is a study conducted by Hamed Nazim Mamache, Gabin Mazue, Osama Rashid, Gewu Bu, and Maria Potop-Butucaru from Seoul, Korea in 2022.

Although this study was conducted to suggest that there are a variety of applications for blockchain technologies, ranging from banking to networking. IOTA blockchain is among the most prominent blockchains designed specifically for IoT environments. In this paper, we investigate the convergence of two IOTA-proposed Consensus: Fast Probabilistic Consensus and Cellular Consensus, when run atop different topologies.

Their research demonstrates that the design of IoT-specific blockchains remains an open research problem and provides design suggestions. Our findings validated the motivation of the IOTA foundation, which is developing a comprehensive version of consensus, Coordicide, for the new IOTA, while viewing these two as constituents. The first being, Fast Probabilistic Consensus (FPC) and the second being Cellular Consensus(CC).

In the method, we are using distributed ledger technology (DLT) that is Tangle which maintains the integrity of the data during the exchange or transfer of data between entities.

In order to overcome this difficulty, the article analysed a proposed blockchain architecture that attempts to achieve the ideal data transmission mechanism[3].

**D.** Towards a blockchain powered Iot data marketplace

This study has been conducted by Pooja Gupta, Volkan Dedeoglu, Salil S. Kanhere and Raja Jurdak from Bangalore in 2021.

The unprecedented rate of IoT adoption presents device owners with an opportunity to sell their IoT data to interested parties. A blockchain-enabled data marketplace can democratise the trading of private IoT data by allowing data owners to decide what and with whom they wish to share their data.

However, certain characteristics of IoT make it challenging to trade generated data on conventional centralised markets. This research focuses on developing a marketplace framework to address design challenges posed by IoT characteristics, including limited resource and computational capabilities, mobility, data privacy, and reselling issues.

They proposed the result that declare propose a three-tiered framework for addressing these challenges from an elemental, functional, and managerial standpoint.

Existing approaches do not account for the IoT characteristics and design challenges outlined previously. Prior to realising the potential of blockchain for IoT data marketplaces, scaling issues surrounding the listing and matching of offers and queries must be resolved[4].

**E.** Towards a Decentralised Data Marketplace for Smart Cities

This study has been conducted by G. S. Ramachandran, R. Radhakrishnan and B. Krishnamachari in 2018.

The interference we get from the above mentioned conference paper is to look at how blockchain and other distributed ledger technologies can be used to create a decentralised data marketplace. Also to consider the potential benefits of such a decentralised architecture, identify various elements that such a decentralised marketplace should have, and demonstrate how they could be potentially integrated into a comprehensive solution[5].

**F.** Towards a blockchain powered IoT data marketplace

This study has been conducted by Pooja Gupta, Volkan Dedeoglu, Salil S. Kanhere and Raja Jurdak from Bangalore in 2021.

The exceptional rate of IoT adoption provides an opportunity for device owners to exchange their IoT data with interested purchasers. A blockchain-enabled data marketplace can democratise the trade of private IoT data by allowing data owners to select what they wish to share and with whom. However, several characteristics of IoT make it impossible to exchange produced data in traditional centralised marketplaces. This study focuses on establishing a marketplace framework to handle design constraints imposed by IoT features such as limited resource and computing capabilities, mobility, data protection, and resale concerns. We offer a three-tiered paradigm for efficiently addressing these difficulties from elemental, functional, and management perspectives[6].

**G.** Trust Modelling for Blockchain-Based Wearable Data Market

This study has been conducted by M. J. M. Chowdhury in 2019.

The study encourages the use IoT smart devices to continuously produce physiological data that can provide individuals critical information about their daily routine or fitness level in combination with their smartphones without requiring manual calculations or maintaining log-books.

The framework suggested will help individuals to sharing the personal data with prospective researchers without the risk of compromising privacy or security[7].

**H.** Blockchain technology: Is it hype or real in the construction industry?

The above mentioned paper was written by Srinath Perera, Samudaya Nanayakkara, M.N.N. Rodrigo, Sepani Senaratne, Ralf Weinand in 2020.

It promotes the use of blockchain technology in the current market in various sectors such as FinTech, IoT and Big Data.

The purpose of this paper was to objectively examine the application potential of blockchains in construction through use case analysis and a complete literature research to determine if it is pure hype or genuine. The investigation indicated that blockchain has a genuine potential in the construction sector due to the exponential usage of blockchain, investments engaged, and a number of start-up enterprises contributing to Industry 4.0[8].

**I.** A Scalable, Standards-Based Approach for IoT Data Sharing and Ecosystem Monetization

The study was made by K. Figueredo, D. Seed and C. Wang in 2022 to enhance the method of data sharing in IoT devices.

The abive mentioned study provides us an insight on the IoT data marketplace architecture and shows how it may be used in smart city and intelligent transportation system implementations. It emphasises the usage of oneM2M, an open standard for the middleware layer in the Internet of Things technological stack. The middleware capabilities of oneM2M are located between IoT devices and communication networks, as well as the AI/ML and IoT applications that consume IoT data. The planned IoT data marketplace also includes capabilities such as licencing, use tracking, and safe data exchange[9].

**J.** Monetization using Blockchains for IoT Data Marketplace

The above paper was written by W. Badreddine, K. Zhang and C. Talhi in 2020 to enhance the use of blockchain and cryptocurrency for the purpose of monetization.

The study tries to overcome the concerns of the trade-off between the overhead of tracking IoT data on a blockchain vs. the accuracy of the monetization for data producers and consumersby implementing a dependable and transparent monetization system based on Distributed Ledger Technology (DLT) and smart contracts. It suggest three monetization strategies and show how the overhead of tracking IoT data on a blockchain compares to the accuracy of monetization for data producers and users. It specifically provide a Bloom filter-based solution for effective data interchange verification using Ethereum and solidity[10].

**K.** An Overview of Blockchain Technology: Architecture, Consensus, and Future Trends

The above paper was written by Z. Zheng, S. Xie, H.N. Dai, X. Chen, H. Wang. It recognises the unequivocally use of blockchain in current tech-trend.

It shows a comprehensive overview on blockchain technology and also it talks about the security challenges that are found while using the technology.

The paper also compares some common consensus methods used in various blockchains. In addition, technical obstacles and recent breakthroughs[11].

**L.** HyperChannel: A Secure Layer-2 Payment Network for Large-Scale IoT Ecosystem

The Above study was conducted by Q. Wang, C. Zhang, L. Wei and Y. Xie.

This study describes a revolutionary distributed layer-2 payment network created exclusively for the IoT environment that safely offloads transaction processing to a collection of Intel Software Guard Extensions (SGXs) controlled by for-profit selfish third parties[12].

## IV. IMPLEMENTATION

The goal is to establish a cross-technology platform integrated with orchestration tools using a suitable blockchain architecture.

Deployed sensors collect the data and send it to the portal, where the similar data get stored in a cluster for future use. These sensors are connected to the database from which data fetches and the user gets authenticated. User verifies its identity through the web portal and get the access of the network. Figure 1, shows the step vise activities performed to build a blockchain based network. A web based platform called Datchain will be made for the user's interaction, which incorporates the database connected. After user gets authenticated, a transaction is published to the blockchain, a new block is created with the user's data and is connected to the network.

Figure 1 - Activity Diagram

A. Blockchain and smart contracts

The application under the Lime light utilises Ethereum BC because it is maintained by a big community, has access to a large collection of libraries, and has a built-in cryptocurrency that enables the development of the payment system without any complications. Each Data MP participant is required to have an Ethereum account. Each action (such as sign in/up, sensor registration, and data stream subscription) is recorded by deploying the associated SCs.B. Application Architecture.

In Figure 2, Hardware sensors that have been deployed will acquire real-time data and transfer it to the front-end through an Application Programming Interface (API).

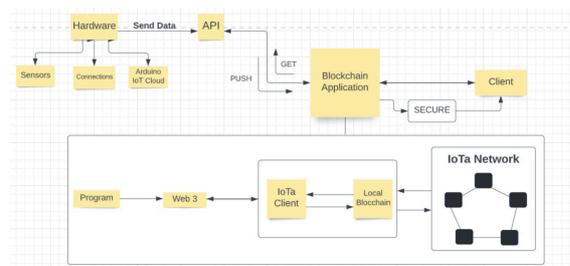

Figure 2 - Sequence Diagram

Our Blockchain-connected platform application leverages Distributed Ledger Technology (DLT) for data encryption and database storage. Three phases make up Blockchain Application: Our programme was initially delivered to Web 3 technology, the third generation in the progression of web technologies. The data is then moved to the subsequent phase, during which IOTA Client and Local Blockchain are interconnected for encryption purposes. The final phase involves the incorporation of data into the larger IOTA network data-marketplace.

## V. SUMMARY

This paper introduces the platform, a secure IoT data market that is modular and built on a microservices-based architecture. The application's objectives were to allow data users in the fields of healthcare, academia, and social studies and many more to transmit a large number of data quickly and securely, as well as to create a secure and encrypted location to store data obtained from IoT sensors.

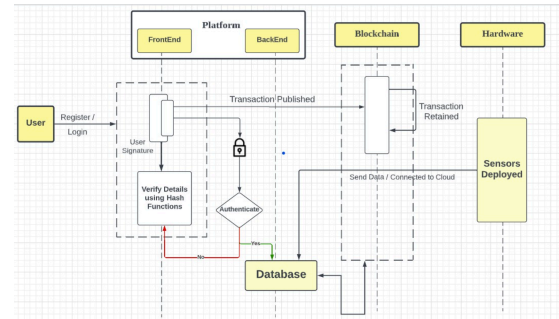

It is intended to create and implement techniques for safe data transmission as well as prevent data buyers from reselling the data once they have it.